# Comment on 'Detection of small comets with a ground-based telescope' by Frank and Sigwarth

R. L. Mutel

Department of Physics and Astronomy, University of Iowa, Iowa City, Iowa, USA

J. D. Fix

Department of Physics, University of Alabama in Huntsville, Huntsville, Alabama, USA

## 1. Introduction

*Frank and Sigwarth* [2001, hereafter referred to as FS] report on an optical search for small comets in which they report detection of nine faint trails. FS claim that these trails are produced by low-mass, low-albedo objects whose inferred number density is consistent with that predicted by the small comet hypothesis. The images used for the search were the same as those obtained in a previous unsuccessful optical search for small comets reported by *Mutel and Fix* [2000, hereafter referred to as MF]. The results of these two independent analyses of the same image dataset are not formally in disagreement, since all detections reported by FS are fainter than the minimum detectable magnitude limit (16.5) determined by MF. However, since the conclusions are quite different, we have independently re-analyzed the original search images[1] for evidence of the faint detections reported by FS. In particular, we have carefully examined whether the three putative trails whose positions are available satisfy several necessary criteria for a candidate grouping of excess-count pixels to be caused by a celestial object.

## 2. Historical Note on Data Sample

Since the detections reported by FS were made using the same images obtained, analyzed, and published by MF, it is curious that FS do not mention this fact. In order to make the historical record clear, we briefly summarize the events surrounding the optical search project beginning in the Fall of 1998. In August 1998, Frank and Sigwarth suggested to one of us (RLM) that the 0.5 m telescope at the Iowa Robotic Observatory would be a suitable instrument for a new optical search for small comets. They suggested using a form of the 'skeet shoot' method originally used by *Yeates* [1989] and by *Frank et al.* [1990] in previous optical searches. After several discussions, we agreed that we would conduct the search completely independent of Frank and his group, including developing the specialized tracking and camera software required for the search, and would conduct the necessary observations for a period of eight months. Neither Frank nor any of his group ever participated in any of the observations, although Frank shared the cost of an equipment repair required for the search. At the end of the search period, we analyzed the observations and reported our results in MF. At the outset, we also agreed that we would share all images with Frank's group and that they would be free to conduct an independent analysis. In the event, this is exactly what transpired: The detections reported in FS are the result of that independent analysis.

In contrast with FS, MF reported that after careful visual inspection of 2,713 search images, no trails could be detected to a limiting magnitude of 16.5 (120 pixel trails). We concluded that if a population of small comets existed, the number density must be less than 5% of the number density of small comets derived by *Frank et al.* [1990]. *Frank and Sigwarth* [2000] subsequently challenged this conclusion based on a disagreement concerning the calculation of the effective search volume. *Mutel et al.* [2001] replied to the objections of *Frank and Sigwarth* [200] by generalizing the search volume calculation to include orbits with a range of perihelia and ecliptic inclinations. We found that relaxing the assumption that small comets orbits lie close to the ecliptic reduces the expected number of detections. However, the expected number was sufficiently large for any plausible range of orbital parameters that the lack of *any* detections meant that the conclusions of MF were not substantially altered.

---

[1] The original search images (FITS format) are available via Web browser or via Anonymous FTP from ftp://ftp.agu.org, directory "apend" (Username = "anonymous", Password = "guest"); subdirectories in the ftp site are arranged by paper number. Information on searching and submitting electronic supplements is found at http://www.agu.org/pubs/esupp_about.html

## 2. Independent Analysis of Frank & Sigwarth Detections

The detection limit used by MF was based on careful calibration of the visual image inspection technique using simulated comet trails (see MF section 3.2). The nine detections reported by FS are all fainter than the detection limit of MF, and are therefore not in direct conflict with our null result. However, since we had access to the original images, we thought it useful to conduct an independent analysis of the FS detections. Unfortunately, we were unable to analyze all nine claimed detections, since repeated requests for the exact trail coordinates in each image have gone unanswered. However, FS illustrates trails for two of the detections; another trail position was obtained from an illustration in a previous talk by Frank on small comets given in February 1999 [*Frank* 1999]. We carefully examined all six remaining original images with candidate trails listed by FS but were unable to find detectable trail signatures on any of the images. In this note we report on analysis of the three trails with known positions.

**Tests for trail validity**

Reliable detection of very faint trails caused by celestial objects in CCD images is a difficult challenge, since many effects can cause artifacts similar to the sought-after trail [*Schildknecht et al.*, 1995]. The statistical significance of trail detection, even with a predetermined location, can only be calculated if the effective number of degrees of freedom for the event in question is known. This is a poorly constrained problem, since both the trail length and orientation are free parameters. In addition, the presence of star trails, CCD sensor blemishes, clusters of hot pixels, variable seeing, and other effects complicate the calculation. Fortunately, a number of plausibility checks can be applied which are very straightforward and avoid these problematic statistical considerations. We tested the validity of candidate trails using the following criteria:

1. The point spread function (width) of a candidate trail must be the same as that of the star trails on the same image within measurement uncertainty.

2. The candidate trail must have the same modulation pattern as the camera shuttering (either a 2:1 or 2:1:1 pattern depending on the observing date; see MF).

3. The trail pattern must be collinear, or if not collinear due to imperfect tracking, it must deviate from collinearity in the same fashion as the star trails on the image.

4. Candidate trails oriented parallel to star trails must be carefully checked to ensure that they are not scintillation-enhanced segments of very faint star trails.

**Analysis of candidate trail 1**

Candidate trail 1 was found on a 40 second exposure obtained on 19 October 1998 starting at 10:43:31 UT. The camera was shuttered using a 2:1 modulation scheme (20 sec open, 10 sec closed, 10 sec open). It is listed as the first detection in FS but not illustrated. However, the trail was shown in an illustration during a public lecture by L. A. Frank [*Frank* 1999] so its position on the image could be determined. In Table 1 of FS, the 'full width' of the trail is listed as 8 pixels (9.9 arcsec). This width is used by FS to sum pixels in order to calculate the statistical significance of the trail. However, this width cannot be correct, since it far exceeds the observed width of the star trails in this image. Fig. 1a shows a cross-sectional profile of a representative star trail in the image. The solid line is a Gaussian function fitted to the profile. The full-width at half-maximum (FWHM) of the Gaussian function is 3.7 arcsec (3.0 pixels). The dashed line is a Gaussian function with an 8 pixel FWHM as stated for the claimed detected trail in FS. The dashed line is incompatible with the measured star trail width and therefore violates the first criterion. Using a box width of three pixels and the claimed trail length of 32 pixels, we were unable to reproduce the 'segment responses' given in Table 1 of FS. Indeed, the entire set of nine 'detections' fails the first criterion: We have compared the measured full width at half maximum (FWHM) of representative star trails for each of the nine images with the 'full widths' listed reported in Table 3 and find no correlation with the measured FWHM's.

The use of an incorrect trail width also affects the claimed object's apparent magnitude since the summed area is incorrect. This is illustrated in Fig. 1b, which shows a section of the trail 1 image. The candidate trail is outlined in the dashed box, following the illustration in *Frank* [1999]. The box at right shows three synthetic trails superposed in the image (see MF for a discussion of synthetic trails) using the same point-spread function as observed for the star trails. Comparing the synthetic trails to the candidate trail, it is clear that a trail with total magnitude 17.0 (as claimed in Table 1 of FS) is much brighter than any putative trail in the dashed box.

### Analysis of candidate trail 5

Fig. 2 shows candidate trail 5 in the original search image taken on January 21, 1999 at 06:22 UT. This trail is parallel to the star trails in this image. Hence, we investigated whether the detection may be part of a faint star trail which was just below the visual detection limit except during short periods when varying clouds or haze may cause intensity fluctuations. To increase the signal-to-noise ratio of the region, we aligned and median-averaged the previous five search images (from 06:13 UT), all of which contained the region of interest. As figure 2a clearly shows, there is a faint star trail exactly aligned with the detection trail. By aligning the search image with the Digitized Sky Survey, it is possible to identify the star trails with individual stars. There are actually two stars with overlapping trails which are responsible for the trail which lies along the small comet detection trail: GSC 1394-0792 (15.2) and USNO-0000363 (16.0). For reference, the brightest trail is the 8th magnitude bright star SAO 98249. We conclude that candidate trail 5 is very likely a segment of a very faint star trail and is not a small comet detection.

### Analysis of candidate trail 8

Candidate trail 8 is located on an image taken on February 6, 1999 at 05:55 UT. This image was taken after we changed the shutter modulation scheme from 2:1 to 2:1:1 (20s open, 10s closed, 10s open, 10s closed, 10s open). This was done to better discriminate against false positive detections. Fig. 3 shows that the trail fails criterion 3, collinearity of trail segments. Figure 3c is an enlargement of the candidate trail with boxes indicating the expected length and width as given in FS Table 3. Note that the long (20 sec) segment is not aligned with the two short segments. This is noted in FS, who assert that the 'displacement is about 1 to 2 pixels and is due to winds and the responses to the motor drives'. Since these effects would affect all trails in the image, this hypothesis can be tested by examining the collinearity of long segments of star trails in the same image. An example is shown in Fig. 3a, for which we have superposed a reference line one pixel wide. Any telescope tracking error will cause identical deviations in star trails and 'detection' trails, both in magnitude and time. The 'detection' trail deviates from collinearity in its first third (longest segment) by 2-4 pixels (Fig. 3c). However, the star trail shows no such corresponding shift. It is collinear (±0.5 pixels) everywhere along its length, and hence disproves the assertion of FS.

We also note that although it is possible for strong winds to affect telescope pointing, we have independent evidence that this was not the case for this image. Each image header in the search contains current weather information acquired from an observatory weather station during that exposure. The wind speed at the time of the image (and indeed for several images before and afterward) was zero miles per hour.

## 2. Expected number of detections as a function of apparent magnitude

An independent test of the small comet hypothesis is whether the integral number density of detections versus magnitude is consistent with the model distribution published in previous small comet papers (e.g. *Frank and Sigwarth* 1999), and illustrated in Figure 7 of FS. In the visual magnitude range fainter than 18.4 (at a range 137,000 km), the model distribution is approximately a power-law of the form

$$N(V) = N_0 \cdot 10^{\alpha \cdot (V - V_0)} \qquad (1)$$

where $N_0 = 3 \cdot 10^{-20}$ m$^{-3}$, $V_0 = 18.4$, and $\alpha = 1.7$. The expected number of detections in the IRO survey sample is given by the product of the number density and the sample volume. *Mutel, Gayley, and Fix* [2001] have evaluated the IRO small comet search sample volume for a variety of possible small comet orbital parameters. For an assumed uniform distribution in inclination angle between 0° and 17° and in semi-major axis between 0.8 AU and 1.0 AU, the sample volume per image is $7 \cdot 10^{16}$ m$^3$. FS used 1,500 images of the IRO search in their analysis so their total search volume is $1 \cdot 10^{20}$ m$^3$. We note that this sample volume only counts objects at low ecliptic inclination (e.g., $|i| < 1.5°$ at $a = 0.9$ AU), and therefore takes into account the undetectability of objects with large velocities perpendicular to the ecliptic, as discussed in FS.

In Fig. 4 we have plotted the expected and observed integral number density as a function of detection magnitude (normalized to 137,000 km). It is clear that the observed distribution strongly disagrees with the predicted distribution both in scale and in form. For example, the integral counts are a factor of 50 lower than expected for V< 19.5. Also, the shape of the observed sample is much flatter than the expected distribution, with a power-law exponent close to $\alpha = 0.5$ (Fig. 3 dotted line). We conclude that even if the claimed detections are real, the observed integral counts reported in FS are inconsistent with the model distribution of the small comet hypothesis.

## 3. Summary and Conclusions

We have examined three of the nine claimed trail detections reported in FS using the original IRO search images. While we have not evaluated the formal statistical significance of faint detections, all three claimed detections fail one or more independent criteria required for a valid detection. In addition, both the level and shape of the claimed integral detection rate versus magnitude are in strong disagreement with the integral number density of the small comets hypothesis. We conclude that unless the remaining claimed detections can prove otherwise, the lack of detections after 'careful examination' of the 1,500 search images described in FS provides the most compelling evidence yet published against the small comet hypothesis.

Acknowledgements

The Digitized Sky Surveys were produced at the Space Telescope Science Institute under U.S. Government grant NAG W-2166. The images of these surveys are based on photographic data obtained using the Oschin Schmidt Telescope on Palomar Mountain and the UK Schmidt Telescope.

## References

Frank, L. A., J. B. Sigwarth, and C. M. Yeates, A search for small solar-system bodies near the Earth using a ground-based telescope: Technique and observations, *Astron. Astrophys.*, *228*, 522-530, 1990.
Frank, L. A., Small comet and our origins: The agony and the ecstasy of the scientific debate, *University of Iowa Presidential Lecture*, 1999 (available at web site: http://smallcomets.physics.uiowa.edu/pdf/feb99_pres_lecture.pdf.
Frank, L. A. and J. B. Sigwarth, Detection of small comets with a ground-based telescope, *J. Geophys. Res. 106*, 3665-3683, 2000.
Frank, L. A. and J. B. Sigwarth, Comment on "An optical search for small comets" by R. L. Mutel and J. D. Fix, *J. Geophys. Res.*, *106*, 24,857-24,863, 2001.
Mutel, R. L., and J. D. Fix, An optical search for small comets, *J. Geophys. Res.*, *105*, 24,907-24, 915, 2000.
Mutel, R. L., K. G. Gayley, and J. D. Fix, Reply, *J. Geophys. Res. 106*, 24,863-24,868, 2001.
Schildknecht, T., U. Hugentobler, and A. Verdun, Algorithms for ground-based optical detection of space debris, *Adv. Sp. Sci. Res.*, 16, 47,50, 1995.
Yeates, C. M., Initial findings from a telescopic search for small comets near Earth, *Planet. Space. Sci. 37*, 1185-1196, 1989.

____________

J. D. Fix, Department of Physics, University of Alabama in Huntsville, Huntsville, AL 35899, USA. (fixj@uah.edu)
R. L. Mutel, Department of Physics and Astronomy, University of Iowa, Van Allen Hall, Iowa City, IA 52242-1479, USA. (robert-mutel@uiowa.edu)

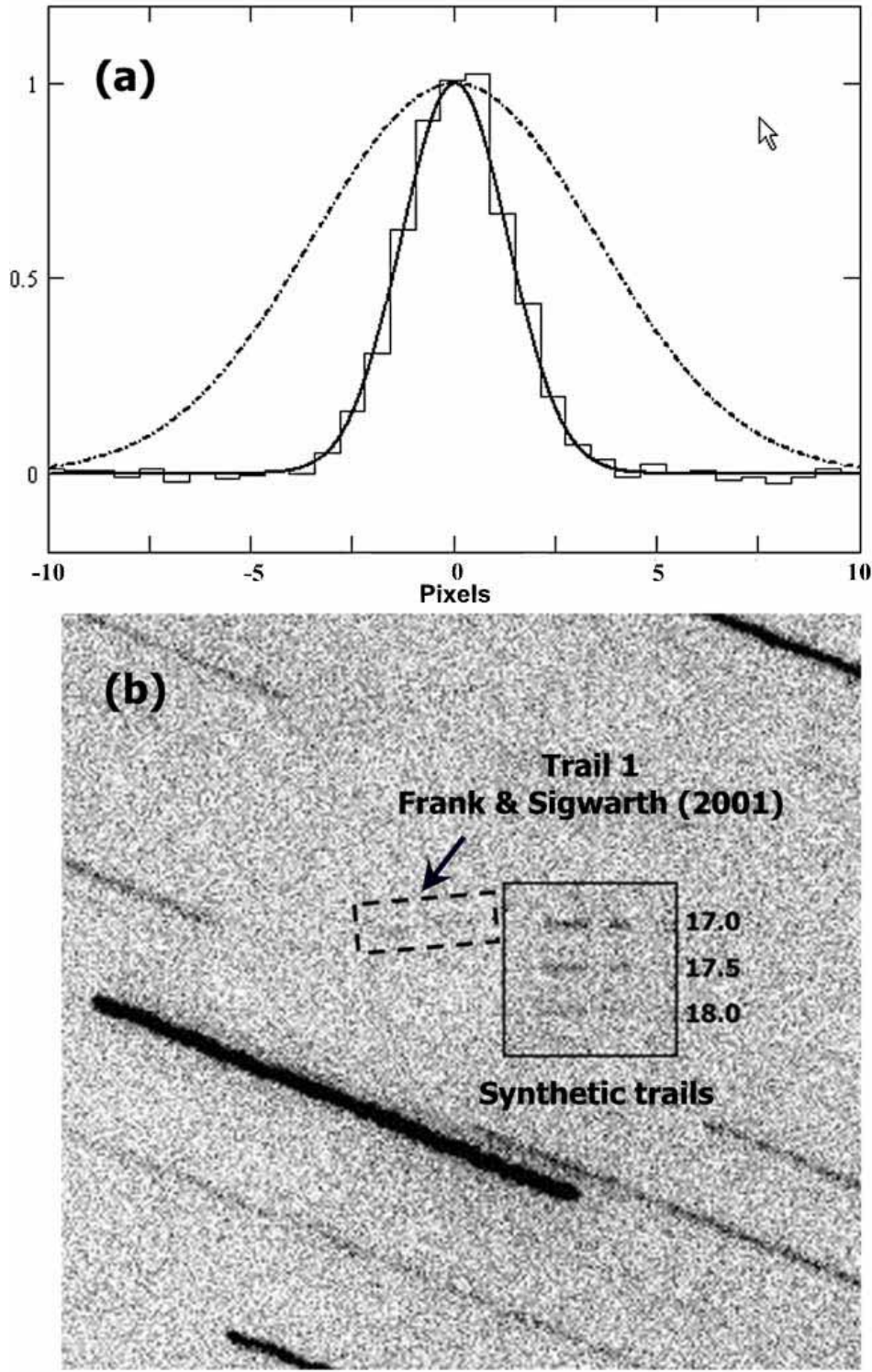

Figure 1. *(a)* Normalized one-dimensional profile transverse to a representative star trail on original IRO search image of 19 October 1998 at 10:43:01 UT. The solid line shows a best-fit Gaussian profile with full-width at half maximum (FWHM) equal to 3.0 pixels (3.7 arcsec). The dashed line shows a gaussian profile with FWHM eight pixels (9.9 arcsec) as claimed by FS for candidate trail 1. This width is clearly much larger than the width of the star trail, violating detection criterion 1. *(b)* Section of image from 19 Oct 1998 with candidate trail outlined (dashed box). The solid box shows synthetic trails superposed on the image with carefully calibrated magnitudes. The trails are the same length and shutter modulation as the FS candidate trail (total length 32 pixels), but have a FWHM equal to the star trails. The claimed apparent magnitude of 17.0 for trail 1 (FS Table 3) is obviously incorrect, probably a result of using the wrong trail width.

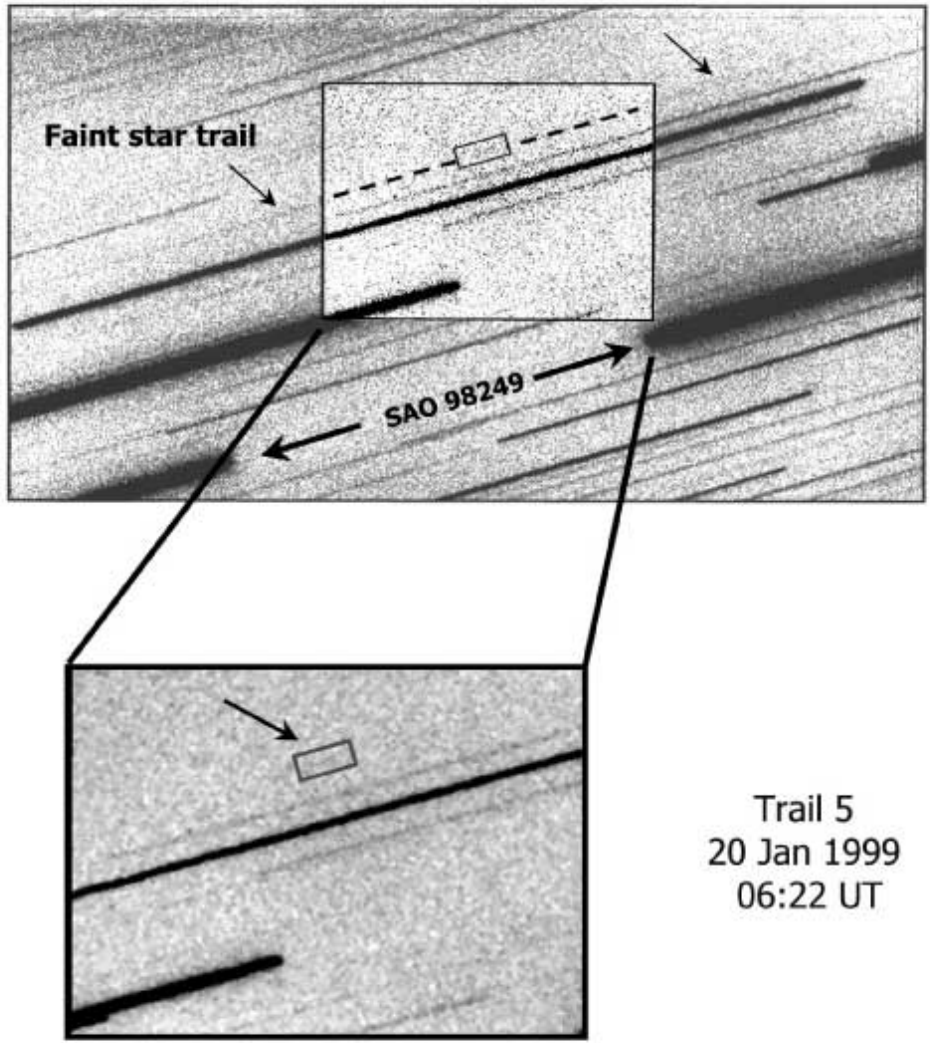

Figure 2. Detection trail 5 (20 January 1999 at 06:22 UT) shown on original search image (below), and superposed on an aligned and median-stacked sequence of images taken between 06:13 and 06:21 UT (above). Note that the detected trail is exactly aligned with a faint star trail. See text for details.

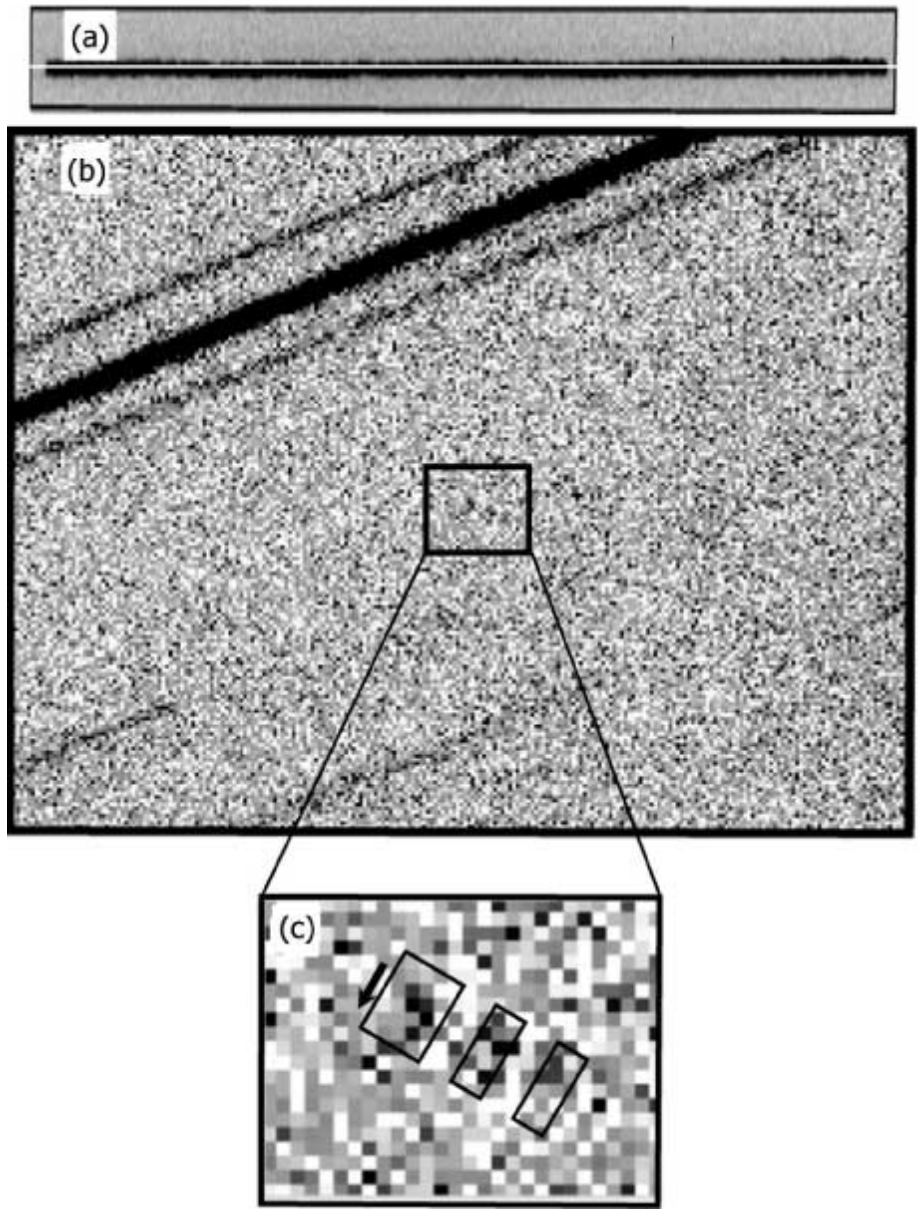

Figure 3. *(a)* First (20 sec) section of a bright star trail from search image containing detection trail 8 (16 February 1999 at 05:55:40 UT) with 1-pixel wide white reference line superposed. The star trail is straight within one pixel and is not consistent with Frank & Sigwarth's claim that the 20 sec section of the detection trail is displaced 'due to winds and the response of the motor drives'. *(b)* Portion of the original detection image; (c) magnified section of image showing detection trail with superposed boxes illustrating the expected 2:1:1 shutter modulation. The boxes have a total trail length (15 pixels) and width (6 pixels) identical to the claimed detection of FS (Table 3). The arrow indicates apparent shift of 20 sec (4 pixel) section (leftmost box).

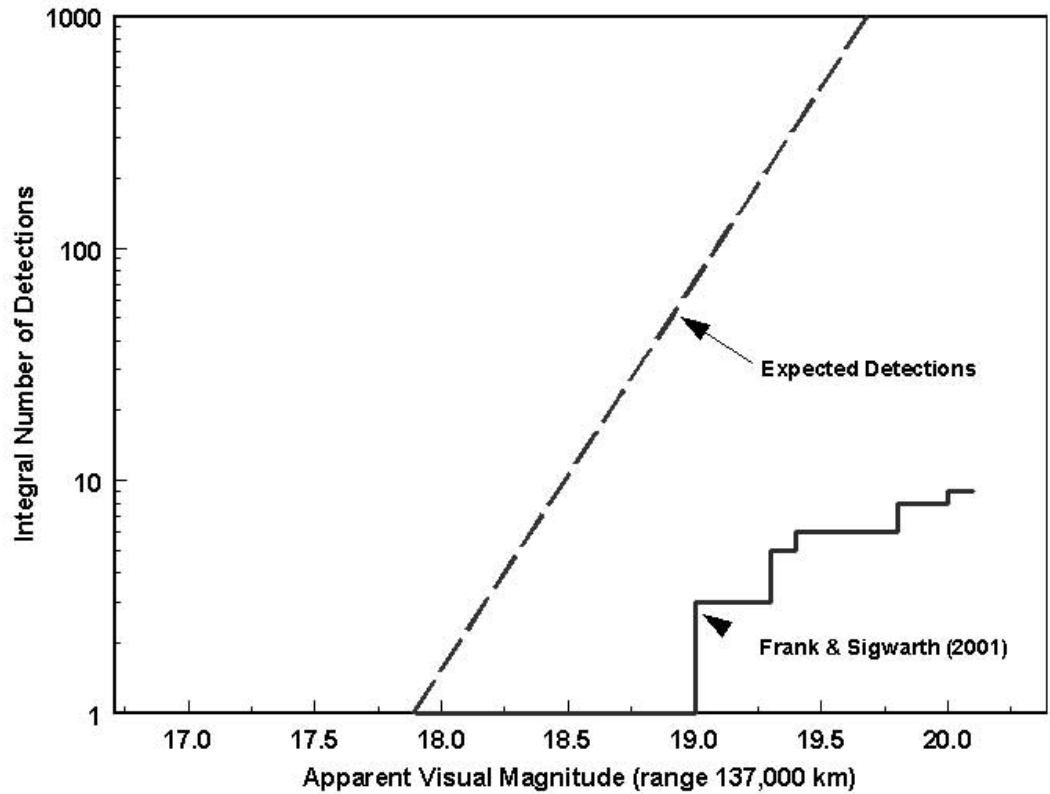

Figure 4. Comparison of expected versus observed integral small comet number density versus magnitude normalized to 137,000 km range. The model distribution is based on a smooth power-law fit to the integral number density distribution of Fig. 7 of FS with the sample volume calculated by *Mutel et al.* [2001]. The observed detection rate is much lower than expected. Furthermore, the slope is much flatter, and inconsistent with the expected distribution.